\documentclass[
aps,nofootinbib,
showpacs,showkeys,preprint
tightenlines,preprintnumbers,] {revtex4}

\usepackage{epsf,epsfig,subfigure,graphicx,amsmath,amssymb 
}
\usepackage{color}
\newcommand{\dis}[1]{\begin{equation}\begin{split}#1\end{split}\end{equation}}
\newcommand{\ie}{{\it i.e.~}}
\newcommand{\etal}{{\it et al.\,}}

\newcommand{\gev}{\,\textrm{GeV}}

\newcommand{\eV}{\,\mathrm{eV}}
\newcommand{\Mp}{M_{\rm P}}

\def\sw0{{$\sin^2\theta_W^0$}}

\def\E6{{\rm E_6}}

\def\EE8{{\rm E_8\times E_8'}}

\begin{document}

\draft

\title{\Large\bf  Gravity wave and model-independent axion}

\author{Jihn E.  Kim}
\address
{  Center for Axion and Precision Physics Research (CAPP, IBS),
  291 Daehakro, Yuseong-Gu, Daejeon 34141, Republic of Korea, and \\
Department of Physics, Kyung Hee University, 26 Gyungheedaero, Dongdaemun-Gu, Seoul 02447, Republic of Korea, and \\
Department of Physics and Astronomy, Seoul National University, 1 Gwanakro, Gwanak-Gu, Seoul 08826, Republic of Korea 
} 
\begin{abstract} 
In this short comment, we notice that the model-independent axion contribution to the graviton mass at just outside the Schwarzschild radius is completely negligible in GW150914. The model-independent axion contribution to the graviton mass at the order $10^{-22}\,$eV might be possible for merger of black holes of mass of order $2\times 10^{14}\,$kg.

\end{abstract}
\maketitle


\section{Introduction}\label{Introduction}

Two black holes merging into one was observed on September 14, 2015 \cite{GW1}.
Even though it is a detection of the classical gravity wave(GW), its implication can be into particle physics realm in the final stage of the merging process. Indeed, it has been shown that massive gravitons could have invalidated their interpretation of the event by the GW if the graviton mass is greater than $1.2\times 10^{-22\,} \eV$ \cite{GW2}. 
In this paper, a possible limit on the graviton mass from the ``invisible'' axion is studied for a static black hole (not considering the details of merging process) if it is arising from string theory.

To see the massive graviton, one usually linearizes the gravity around the flat Minkowski background $\eta_{\mu\nu}$,
\dis{
g_{\mu\nu}=\eta_{\mu\nu}+\frac{1}{\Mp}h_{\mu\nu},\label{eq:linearization}
}
where $\Mp\simeq 2.4\times 10^{18\,}\gev$
is the reduced Planck mass,
and obtains a kinetic term for $h_{\mu\nu}$, consistent with the deffeomorphism invariance, and the coupling to matter of the form $h^{\mu\nu}T_{\mu\nu}$ where $T_{\mu\nu}$ is the stress–energy tensor. Phenomenologically, massive graviton couplings have the forms
\dis{
{\cal L}_{\rm int}=ah_{\mu\nu}h^{\mu\nu}+
b(\eta^{\mu\nu}h_{\mu\nu})^2,
}
for which the expected five polarizations of a massive graviton is obtained for $a=-b$ as shown by Fierz and Pauli(FP) \cite{FP39}. Any other choice will lead to a ghost degree of freedom.  

With the FP massive gravity, there are three more degrees compared to two degrees of massless gravity theory. Van Dam and Veltman(VV) noticed a smaller light bending in the  FP massive gravity theory compared to the bending in the massless gravity theory \cite{VanVel70}. However, Vainstein noticed that the VV discrepancy is an artifact of linearized gravity, and   showed  that a scale dependent forces recovers no discrepancy \cite{Vainshtein72}. But, there results an effective Vainstein radius. So, to an observer far outside the Vainstein radius, the description by the linearized gravity  seems enough.

With the FP condition $b=-a$,
\dis{
\frac{\delta \cal L}{\delta h_{\mu\nu}}=2a h_{\mu\nu}+\cdots
}
where $\cdots$ has nothing to do with the mass.
The equation of motion is
\dis{
\square^2 h_{\mu\nu}=-2a\,h_{\mu\nu},
}
and $\frac12 a$ is the graviton mass $m_g^2$.

\section{Model-independent axion contribution to the graviton mass}\label{Pseudoscalar}

However,  light particles (except massive graviton) cannot contribute to the radiation in the event of GW150914. The main reason is that the effective coupling of light particles to curvature is higher order. If they couple at O($R$), one removes its coupling by going into the Einstein frame. So, if it is effective,  one considers O($R^2$) couplings.  Let us illustrate this effect in the event of GW150914 for the model-independent(MI) axion.

Let us start with the Einstein equation 
\dis{
R_{\mu\nu}-\frac12 Rg_{\mu\nu}=\frac{1}{\Mp^2}T_{\mu\nu}.
}
With the linearization of (\ref{eq:linearization}), it gives the following equation
\dis{
\square^2 h_{\mu\nu}=\frac{1}{\Mp^2}\frac{\delta T_{\mu'\nu'}}{\delta g_{\mu'\nu'}}h_{\mu\nu}.\label{eq:Tmunutogrmass}
}
So, if $T_{\mu\nu}$ has a non-derivative contribution of $g_{\mu\nu}$, it can contribute to the graviton mass.

Among pseudoscalar fields in the literature, there exists only one such field contributing to the graviton mass in the high curvature circumstance: the model-independent axion, $a_{\rm MI}$, in string compactification \cite{GS84,WittenMI84,ChoiKim85}. It belongs to anti-symmetric tensor field in ten dimension(10D), $B_{MN}\, \{M,N=1,2,\cdots,10\}$. $B_{\mu\nu}\, \{\mu,\nu=1, \cdots,4\}$ is the MI axion. Usually, it is given in the dualized form $\partial_\mu a_{\rm MI}\propto \epsilon_{\mu\nu\rho\sigma}H^{\nu\rho
\sigma}$\cite{WittenMI84}. where $H_{\nu\rho\sigma}$ is the field strength of $B_{\rho\sigma}$.

The ten dimensional(10D) quantum field theory(QFT) with $\EE8$ gauge group has 10D anomalies. But the heterotic string is anomaly free. So, the string theory must have a term which cancels the QFT anomaly in the point particle limit. Indeed, the string theory has the anti-symmetric tensor field $B_{MN}$ which allows such a term, removing the anomaly in the point particle limit.  It is the Green-Schwarz term \cite{GS84}. The field strength of $B_{MN}$ in differential forms, $H$, satisfies 
\dis{
dH=\frac{1}{30}\textrm{Tr\,}F^2_{E_8}+ \frac{1}{30}\textrm{Tr\,}F^2_{E_8'}- \textrm{Tr\,}R^2.
}
Under the general coordinate transformation, the MI-axion coupling behaves like the cosmological constant because its coupling to the metric is only through the overal multiplication $\sqrt{-g}$, viz. the action
\dis{
\propto \int\sqrt{-g}\, d^4x \,
\frac{a}{f_a}\,\epsilon^{\mu\nu\rho\sigma}\left( F_{\mu\nu}F_{\rho\sigma}-R_{\mu\nu}R_{\rho\sigma}  \right).
}
Thus, $T_{\mu\nu}$ involves a coupling of the following form,
 \dis{
 T_{\mu\nu}= -g_{\mu\nu}{\cal L}=g_{\mu\nu}\frac{a}{f_a}(F\tilde{F} -R\tilde{R}) .\label{eq:GSterm}
 }
  In fact, the action contains more terms involving $g_{\mu\nu}$. But, the important one is those leading to large values just outside the blackhole surface. Terms in ${\cal L}$ in the action is just one particle term which can be neglected. The field values at the surface is enhanced if there is a long range force such that all particles inside the blackhole contribute. The obvious one is the gravitational force since all energy inside the blackhole contribute. Also, all electromagnetic charges inside the blackhole contribute to {\bf E} field at the surface. However, it is exponentially suppressed to have a great number of electric charges inside the blackhole. Statistically, the ratio of charges of proton to the number of nucleons inside the blackhole is of order $1/\sqrt{N}$, which is completely negligible for a macroscopic blackhole. Therefore, we consider only the $  R\tilde{R}$ term in Eq. (\ref{eq:GSterm}).
 
Then,  we estimate the MI-axion contribution to the graviton mass as
 \dis{
  \frac{a}{f_a}\frac{R\tilde{R}}{\Mp^2}.\label{eq:MIrad}
 }
 
The LIGO group distinguished the merging process by three stages,  the  inspiral, intermediate, and merger ringdown stages. For our purpose, consideration of inspiral may be enough. For the merger event GW150914, the inspiral process had the oscillation counts from No. 20 to No. 50, \ie 30 oscillations.  These 30 oscillations match their numerical gravity calculation and if any other sources of radiation were present, it might have changed their numerical solution. Thus, they obtained the upper bound on the graviton mass, $m_g<1.2\times 10^{-22\,} \eV$.

If the MI-axion were present, then Eq. (\ref{eq:MIrad}) contributes to the graviton mass, which is time-varying because the axion field is oscillating now. The mass of the MI axion is  for the axion mass in the range $1.6\times 10^{-9} \,\eV$ which corresponds to $f_a\simeq 0.4\times10^{16\,}\gev$ \cite{ChoiKim85,KimPLBNDW16} and the   axion oscillation frequency is of order $2\times 10^{6}\,{\rm s}^{-1}$, which will have many oscillations during one period of the inspiral gravity wave of GW150914. Thus, we average this axion oscillation contributing to the radiated graviton mass  in the inspiral regime, $\frac{1}{50}\,{\rm s}^{-1}-\frac{1}{30}\,{\rm s}^{-1}$.  Even though the axion field oscillate, the radiated-away MI-axion cannot come back and the contribution to the radiated energy is always considered to be positive. So, we take the root mean square of the vacuum axion oscillation, \ie we use $\langle a\rangle=|a_{\rm present~max}|/2$. The value $|a_{\rm present~max}|$ is about $10^{-20}f_a$.\footnote{See, for example, Eq. (10.35) of Ref. \cite{KimPRP87}.} Thus, comparing to the   graviton mass, the radiated energy parameter must satisfy,  
\dis{
  \frac12 10^{-20} \langle R\tilde{R}\rangle\lesssim  \Mp^2\left(1.2\times 10^{-22\,} \eV \right)^2\simeq  10^{-25\,} \gev^4 .\label{eq:OurIneq}
}

Around the Schwarzschild radius $r_{SC}=2M/\Mp^2$, the square of Riemann curvature tensor is $R^{\mu\nu\rho\sigma}R_{\mu\nu\rho\sigma}
=48M^2/\Mp^4r^6$ where $M$ is the blackhole mass and $r$ is the point of interest from the origin. This is for the cosmological constant zero at $r=\infty$. Since we need only the information on the curvature around the BH, we estimate it based on this solution. At $r=r_{SC}=2M/\Mp^2$, $I\equiv R^{\mu\nu\rho\sigma}R_{\mu\nu\rho\sigma} = 3\Mp^8/4M^4 $. Instead of directly calculating $R_{\mu\nu}\tilde{R}^{\mu\nu}$, we use $I$ for an order of magnitude estimate for $|R|^2$ terms around a BH. Since we are interested in the upper bound on $|R|$, this rough substitute will be enough.
 
Since GW150914 starts with two masses $29 M_{\odot}$ and $36 M_{\odot}$, we use $M\simeq 30M_{\odot}$ as a rough estimate. We are interested in the magnitude of $R^2$ just outside the  Schwarzschild radius of GW150914, and will use the  Schwarzschild radius $r=r_{SC}$ for the estimate. For such a blackhole, $I\simeq 0.93\times 10^{-6}\Mp^4(\Mp/ M_{\odot})^4=3.486\times 10^{67\,}\gev^4(\Mp/ M_{\odot})^4$. In this approximation, the left-hand side of Eq. (\ref{eq:OurIneq}) for a BH of $30 M_{\odot}$ mass is
\dis{
1.7\times 10^{47\,}\gev^4\left(\frac{\Mp}{ M_{\odot}}\right)^4= 1.7\times 10^{47}\,[\gev^4]\times 3.4\times 10^{-142} =5.8\times 10^{-95}\,\gev^4
}
which is  certainly smaller than the right-hand side of (\ref{eq:OurIneq}). Thus, the effect of the MI axion to the graviton mass cannot have been detected in GW150914.

It will be of interest to see the size of BHs where the MI axion contribution is comparable to the radiation arising from O($R$) term of  GW150914, $10^{-25\,} \gev^4 $, with $m_g\simeq  10^{-22\,} \eV$.
At the Schwarzschild radius, $I$ is
\dis{
 I=\frac{48M^2}{ \Mp^4 r_{SC}^6}=\frac{3}{4  }\frac{\Mp^{8}}{   M_{\odot}^4} \left(\frac{  M_{\odot}}{M }\right)^4=2.5\times 10^{-142}\Mp^4\left(\frac{  M_{\odot}}{M }\right)^4=0.9\times 10^{-68\,}\gev^4\left(\frac{  M_{\odot}}{M }\right)^4.
}
Approximately, $ 10^{-20}$ times $I$
must be about $10^{-25\,} \gev^4 $:   
$M\simeq 10^{-16}M_{\odot}\simeq  2\times 10^{14}\,$kg, \ie corresponding to the BH mass of $3\times 10^{-11}$ times the Earth mass.
The mass is at the boundary of the primordial black holes studied in \cite{Carr10}.  Black holes with $M<10^{12}\,$kg might have completely radiated away by now. So, the mass region $10^{12}-10^{14\,}$kg range is still possible to be probed.

\acknowledgments{
 I have benefitted from discussions with M. Evans. 
J.E.K. is supported in part by the National Research Foundation (NRF) grant funded by the Korean Government (MEST) (NRF-2015R1D1A1A01058449) and  the IBS (IBS-R017-D1-2016-a00).}

\end{document}